\documentclass[letterpaper, twocolumn, 10pt]{article}
\usepackage{float}
\usepackage{booktabs}
\usepackage{inputenc}
\usepackage[english]{babel}
\usepackage{amsmath}
\usepackage{mathtools, cuted}
\usepackage{amsfonts}
\usepackage{amssymb}
\usepackage{graphicx}
\usepackage{listings}
\usepackage{rotating}
\usepackage{authblk}
\usepackage{placeins} 
\usepackage{hyperref}
\usepackage{cleveref}
\hypersetup{breaklinks=true}
\usepackage{subfig}
\usepackage[margin=1in]{geometry}

\usepackage{multirow}

\begin{document}

\title{A Data Mining Approach for Detecting Collusion in Unproctored Online Exams}
%
%
%
%
%

\author[a]{Janine Langerbein}
\author[a]{Till Massing\thanks{Corresponding author. Email:
till.massing@uni-due.de}}
\author[b]{Jens Klenke}
\author[a]{Natalie Reckmann}
\author[b]{Michael Striewe\thanks{For more information regarding JACK contact this author.}}
\author[b]{Michael Goedicke}
\author[a]{Christoph Hanck}
\affil[a]{\footnotesize Faculty of Business Administration and Economics, University of
Duisburg-Essen, Universit{\"{a}}tsstr.~12, 45117 Essen, Germany}
\affil[b]{\footnotesize paluno - The Ruhr Institute for Software Technology,
University of Duisburg-Essen, Gerlingstr.~16, 45127 Essen, Germany}



\maketitle
\begin{abstract}
Due to the precautionary measures during the COVID-19 pandemic many universities offered unproctored take-home exams. We propose methods to detect potential collusion between students and apply our approach on event log data from take-home exams during the pandemic. We find groups of students with suspiciously similar exams. In addition, we compare our findings to a proctored comparison group. By this, we establish a rule of thumb for evaluating which cases are ``outstandingly similar'', i.e., suspicious cases.
\end{abstract}

%


\section{Introduction}\label{sec:intro}

During the COVID-19 pandemic many universities, e.g., in Germany, were forced to switch to online classes. Moreover, most final exams were held online. In pre-pandemic times, computer-based final exams have already proven their worth, but with the difference that they were proctored in the classroom. During the pandemic this was mostly unfeasible and students had to take the exam from a location of their choice. 

There exists a wide range of supervisory measures for take-home exams. E.g., one could use a video conference software to monitor students. At many universities, however, this is legally prohibited due to data protection regulations. The exams are therefore conducted as open-book exams, i.e., students are allowed to use notes or textbooks. Yet, students must not cooperate with each other. Any form of cooperation or collusion is regarded as attempted cheating.

To our knowledge, it exists no universally-applicable method for proctoring take-home exams. It is therefore hardly feasible to stop students from illegally working together. However, one can attempt to identify colluding students post-exam. The attempt alone could have a deterring effect on students. Research in this area, however, is scarce. \cite{cleophas2021s} present a method for comparing exam event logs to detect collusion. They use a simple distance measure for time series, i.e., the event logs of two different students, to quantify the similarity of these student's exams. Building on this, we propose an alternative distance measure, as well as the use of hierarchical clustering algorithms, to detect groups of potentially colluding students. We find that our method succeeds in finding groups of students with near identical exams. Furthermore, we present an approach to categorise student groups as ``outstandingly similar'', by providing a proctored comparison group. 

The remainder of this paper is organised as follows: \Cref{sec:related} provides a brief overview of related work. \Cref{subsec:data} describes the available data. \Cref{subsec:method} presents our method, including the calculation of the distance matrices. \Cref{sec:analysis} discusses the empirical results. \Cref{sec:conclusion} concludes.

\section{Related Work}\label{sec:related}

Due to the limited relevance of unproctored exams at universities before the pandemic, there exists little research about this topic. Recent work from \cite{cleophas2021s} presents a method for analysing exam event logs for the detection of collusion in unproctored exams. They visually compare the event logs of pairs of students and quantify these by calculating a distance measure. They find some suspicious pairs of students with very similar event logs. Still, the authors remark that these findings might be purely coincidental. We enhance their approach by including a comparison group for drawing the line between ``normal degree of similarity'' and ``outstandingly similar''.

In other contexts, collusion in exams has been a relatively well studied topic. \cite{Hellas_2017,Leinonen_2016} quantify the similarity of programming exams. For this, they calculate distance measures based on student's keyboard patterns. \cite{ihantola_2015} further provide an overview of relevant work in educational data mining in programming exams. Complementary, our work does not focus on keyboard patterns in programming but on the submissions of answers and achieved points in introductory statistics classes. Thus, our calculation of distance measures follows a different approach. 

Furthermore, a major body of related literature focuses on a different methodology. E.g., \cite{bowers,hemming2010online,mccabe2001cheating,olt2002ethics,shon2006college} use surveys or interviews with students to collect data. Due to issues inherent to surveys and interviews, like nonresponse or incorrect responses, there is little knowledge on student collusion based on actual student behaviour. We attempt to bridge this gap by directly using student's exam data. 

Generally, there exists a wide range of proctoring options during take-home exams. \cite{goldberg2021programming, hussein2020evaluation} introduce and compare some of these options. A supervisor could, for example, use video conference software to observe students during the exam. This provides conditions similar to those at classroom exams and thus prevents students from colluding. Such actions have two drawbacks: First, \cite{cluskey2011thwarting} argue that proctoring take-home exams is relatively costly, so that the costs exceed the potential benefits. Second, as mentioned before, most proctoring options are strictly illegal in some countries, e.g., Germany. 

On the other hand, e.g., \cite{miltenburg2019online} advise against unsupervised online exams. They argue that, logically, with no supvervision there is no way to prevent students from colluding during the exam. To date, there are only few studies examining the impact of unattended online examinations on the integrity of students, see, e.g., \cite{manoharan2020upholding}. \cite{harmon2008online} use a regression model that predicts final exam scores to detect collusion in unproctored online exams. Their findings suggest that collusion took place when the final exam was not proctored. \cite{hollister2009proctored} compares the Grade Point Average (GPA) of students who wrote a proctored exam and students who wrote an unproctored exam. There was no evidence of a significant difference in the mean GPA between the two groups, which, however, does not establish that the students did not collaborate illegally. \cite{fask2015integrity} also compare the GPA in a proctored vs. an unproctored online exam and use a regression analysis to measure student collusion. The data used in these studies were final exam scores or the GPA. None of them uses data collected during the exam.

\section{Methodology}\label{sec:methodology}
In the following we give a brief overview about the data used in our analysis. We further describe our approach to build a suitable distance metric. 

\subsection{Data}\label{subsec:data}

The data we use stems from the introductory statistics course ``Descriptive Statistics'' at the Faculty of Business Administration and Economics 
at the University Duisburg-Essen, Germany.\footnote{All personal data was pseudonymised. The chair and the authors have followed the General Data Protection Regulations (GDPR) by the EU as well as national law. Reproducible Code and toy data can be found at \url{https://github.com/Janine-Langerbein/EDM_Detecting_Collusion_Unproctored_Online_Exams}.  For information on access to the actual dataset, please contact the Dean of the Faculty of Business Administration and Economics at the University of Duisburg-Essen (dekanat@wiwi-essen.uni-due.de).} 
The exam of our test group was taken unproctored during the global COVID-19 pandemic in the winter term 2020/21. The exam of our comparison group took place in the winter term 2018/19, i.e., before the pandemic.\footnote{The course is jointly offered by two chairs and therefore held on a rotating basis. Hence, the exam data is only comparable every two years.} It was a proctored exam located in a PC-equipped classroom at the university. Both exams use the e-assessment platform JACK \cite{Schwinning2017}.

Both exams consist mainly of arithmetical problems, where students are expected to submit numerical results. Moreover, there exist some tasks where students are obliged to use the programming language R \cite{R}. The test group also had to answer a short essay task which should contain 4-5 sentences. All but the free-text tasks are evaluated automatically by JACK. The latter is manually graded by the examiner.

During the exam, the students' activities are stored in said event logs. Hence, these contain the exact time for all inputs in all tasks. For all tasks students can change and re-submit their entries. The last submission will be evaluated. For this reason, one task can list multiple events in the event log. 

In addition to the event logs we also use the points achieved per task for our analysis.

\begin{table}
{\footnotesize
	\caption{\label{Tabelle} This table gives summary statistics for all students 			considered in our empirical analysis}
	\begin{center}
	\begin{tabular}[h]{lcccc}
	\toprule
  	Year & Minutes & Points & Subtasks & Students \\ 
  	\midrule
 	\parbox{2cm}{Comparison Group \\ (18/19) \\} & 70      & 60     &  19      & 109 \\
	\parbox{2cm}{Test Group \\ (20/21)}  & 70      & 60     &  17      & 151 \\
	\bottomrule
	\end{tabular}
	\end{center}
}
\end{table}

\Cref{Tabelle} displays the basic data for both exams. Namely, these are the duration and maximum points to achieve, as well as the number of subtasks and participants per group. The wide disparity in student participants between both exams can be explainend by a change in examination regulations. During the COVID-19 pandemic, ergo in the test group, students were allowed to fail exams without any penalties. In order to prevent this from biasing our results, we removed students who attended the exam for only a few minutes and those who achieved merely a fraction of the maximum points.\footnote{We also conducted the analysis without the removal of these students, with no effect but a reduced interpretability of the following clustering algorithms.} We also removed twelve students from the test group who reported internet problems during the exam.

From our perspective, the setup is reasonably comparable in both groups. Although the lecture of the comparison group was held in presence and the lecture of the test group was held online, both groups shared the same content and learning goals. Both times students were given the opportunity to ask questions during the lecture. The amount of those questions remained approximately stable. Due to the sheer size of the course, with more students attending classes than participating in the exam, direct discussions were sparse even pre-pandemic.

\subsection{Model}\label{subsec:method}
We adopt an exploratory approach for finding clusters of students with similar event patterns and points achieved during the exam. For this, we use agglomerative, i.e., bottom-up, hierarchical clustering algorithms. The results are depicted in a dendrogram. We build on previous work by \cite{hastie_09} and \cite{ClusterEvents}.

In general, clustering algorithms attempt to group $N$ objects according to some predefined dissimilarity measure. Those objects have measurements $x_{ij}$ for $i = 1, 2, \ldots, N$, on attributes $j = 1, 2, \ldots, h$. The global pairwise dissimilarity $D(x_i, x_{i'})$, with $x_i$ being $x_{ij}$ over all $j$, between two objects $i$ and $i'$ is defined as
\begin{align}
D(x_i, x_{i'}) = \frac{1}{h} \sum^h_{j = 1} w_j \cdot d_j(x_{ij}, x_{i'j}); \sum^h_{j = 1} w_j = 1, \label{eq:eq1} 
\end{align}
with $d_j(x_{ij}, x_{i'j})$ the pairwise attribute dissimilarity between values of the $j$th attribute and $w_j$ the weight of the attribute. The clustering algorithm therefore takes a distance matrix as input.

In our case, the students are the objects to be clustered, with $N = 151$ students. As attributes we use the dissimilarities in the student's event patterns and the dissimilarities in their points achieved. Both need to be calculated differently, but over all subtasks. Hence, we split $d_j(x_{ij}, x_{i'j})$ into two parts. 

We call the attribute dissimilarity for the points achieved $d_j^P(s_{ij}, s_{i'j})$ with $w_j^P$ its corresponding weight. $s_{ij}$ denotes the points achieved by student $i$ in the $j$th subtask. Since there are $17$ subtasks we obtain a total of $h = 34$ attributes. To receive a dissimilarity measure we calculate the absolute differences
\begin{align}
d_j^P(s_{ij}, s_{i'j}) = |s_{ij} - s_{i'j}|. \label{eq:eq3} 
\end{align}

Next, $d_j^L(v_{ij}, v_{i'j})$ describes the dissimilarities in the event patterns per subtask. To calculate these we divide the examination time into $m = 1, \ldots, K$ intervals of one minute. Since both exams each took 70 minutes, we obtain $K = 70$ intervals. We count each student's answer per interval. The count is denoted with $v_{ijm}$.\footnote{We consider this an enhancement of the distance measure used in \cite{cleophas2021s}, as it enables us to analyse exams with more than one answer per task.} To obtain a pairwise attribute dissimilarity measure for all subtasks, we calculate the Manhattan metric over all counted quantities
\begin{align}
d_j^L(v_{ij}, v_{i'j}) = \sum^{K = 70}_{m = 1} |v_{ijm} - v_{i'jm}|. \label{eq:eq4} 
\end{align}
The corresponding weight is denoted by $w_j^L$. To ensure better transparency, we provide a detailed explanation of each variable in Appendix~\ref{sec:vardesc}.

Finally, we modify \eqref{eq:eq1} so that
\begin{align}
D(s_i, s_{i'}, v_i, v_{i'}) = & \frac{1}{h} \sum^h_{j = 1} \left( w_j^P \cdot d_j^P(s_{ij}, s_{i'j}) \right.  \nonumber  \\
& \left.  + \ w_j^L \cdot d_j^L(v_{ij}, v_{i'j}) \right) \nonumber \\
& \text{with} \sum^h_{j = 1} w_j^P + w_j^L = 1.
 \label{eq:eq2} 
\end{align}

The attribute weights $w_j$ control the influence of each attribute on the global object dissimilarity. If all $34$ attributes are to be weighted equally, each attribute would be assigned a weight of $\frac{1}{34}$. Here, however, we weight the attributes with regard to our research question. Specifically, we observe that students submit entries more often in the case of R-tasks, viz.~subtasks $6a, 6b$ and $6c$. One possible interpretation of this is that students submit their code more often to check its executability. Furthermore, task $7$ demands the answer to be a short text which was corrected manually. This could lead to insufficient comparability between students due to accidental arbitrariness during correction. Based on these aspects, it appears reasonable to reduce the weight of said subtasks.

We further reduce the influence of the points achieved during the exam by decreasing their weight. This follows from the fact that prior to the exam we must define all (partially) correct answers in JACK. In doing so, it is not feasible to anticipate all types of mistakes resulting from, e.g., calculation errors made by students.\footnote{The tasks are randomised, i.e., there exist variations so that sharing exact results is not expedient for the students.} Students might receive no points due to careless mistakes, while still having employed a correct solution strategy. In our view, this might impede the detection of colluding students, e.g., if there exist large differences in points as one student makes more frequent careless mistakes due to the random numbers in the tasks. On this account, we assign smaller weight to the points achieved. For greater clarity, an overview with all exact final weights for all attributes can be found in Appendix~\ref{sec:weights}.\footnote{A robustness check regarding the object weights can be found in Appendix~\ref{sec:robust}. In this analysis, the weights of the objects are equal. The results are basically identical.} 

The influence of each attribute on object dissimilarity further depends on its scale. We therefore normalise each attribute.

\begin{figure*}
\includegraphics[width = \textwidth]{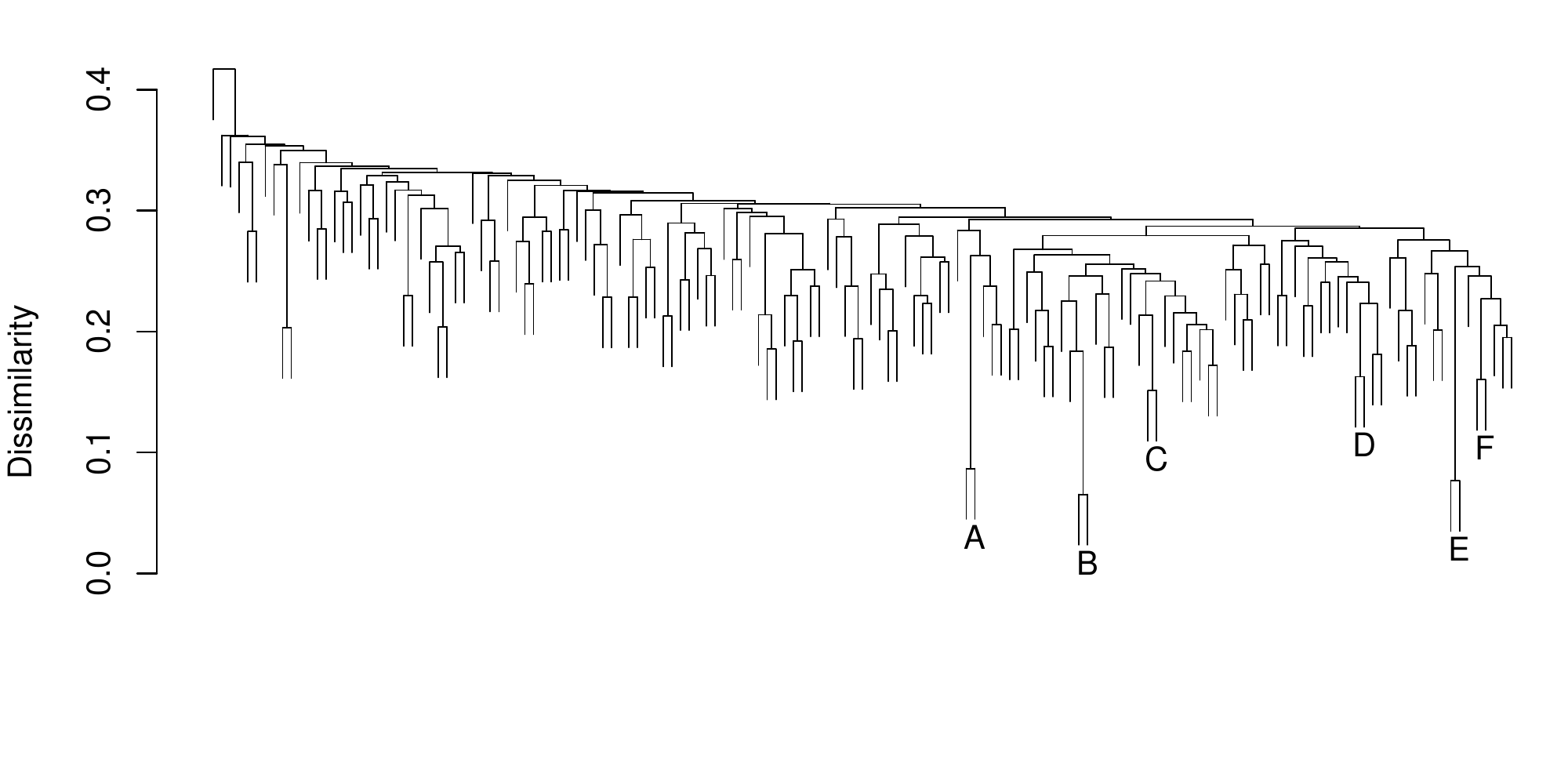}
\caption{Dendrogram produced by average linkage clustering for the test group 2020/21. The dissimilarity of each group's node is displayed on the $y$-axis. Its value corresponds to the dissimilarity of the group's left and right member. A - F denotes the six lowest dissimilarity clusters. Of these especially, clusters A, B, and E are notable.}
\label{fig:tree_cl}
\end{figure*}

From these pairwise object dissimilarities, we create the distance matrices. We then apply agglomerative hierarchical clustering. This builds a hierarchy by merging the most similar pairs of students, viz.~those with the lowest object dissimilarity $D(x_i, x_{i'})$, into a cluster. This is repeated $N - 1$ times, until all students are merged into one single cluster. The merging process is implemented with different linkage methods. These differ in their definition of the shortest distance between clusters. Here, we use single, average and complete linkage. The former merges clusters with the closest minimum distance, the latter uses the closest maximum distance. The average linkage method (here: unweighted pair group method with arithmetic mean) defines the distance between any two clusters as the average distance among all pairs of objects in said clusters.

\section{Empirical results} \label{sec:analysis}
We present the results of the hierarchical clustering algorithms in a dendrogram. This provides a complete visual description of the results from the agglomerative hierarchical clustering algorithm. A dendrogram resembles a tree structure where each object is represented by one leaf. In a bottom-up approach, the objects are merged into groups one by one according to their dissimilarity. Hence, each level of the tree corresponds to one step of the clustering algorithm. The junction of a group is called a node. 

There exist various hierarchical clustering algorithms. Each has a different definition of the distance between groups of observations as a function of the pairwise distances. We calculate the cophenetic correlation coefficent to assess how faithfully each algorithm represents the original structure in the data. 
Appendix~\ref{sec:dendro}, Table~\ref{tab:table_coph_cor_both} gives an overview of the cophenetic correlation coefficient for the different linkage methods for the test and comparison groups. Based on this, we deem average linkage clustering the most suitable algorithm.
 \Cref{fig:tree_cl} shows the corresponding dendrogram. The dissimilarity of each group's node is plotted on the vertical axis. Its value corresponds to the dissimilarity of the group's left and right member.

It is important to note that a dendrogram only gives an indication of clusters which best fit the data. It is up to the analyst to decide which are to be examinated in further detail. 

The dendrogram has a slightly elongated form. Still, compact clusters were produced at medium dissimilarities. This general shape is typical for the underlying algorithm, as average linkage clustering combines the long form of single linkage clustering with the smaller, tighter clusters of complete linkage clustering. Additionally, we observe three notable clusters (A, B and E) which form at a significantly lower height. Each of these three clusters consists of two students. Prima facie, this indicates the absence of collusion in larger groups.

As explained above, the hierarchical algorithm does not cluster the data itself, but imposes a structure according to the students' dissimilarities. There exist various formal methods to decide on an optimal number of clusters given this established hierarchy. Since our primary interest lies in the detection of clusters at low dissimilarities, instead of the general structure of the data, we exemplary investigate the six lowest clusters (A - F) in \Cref{fig:tree_cl}.

\begin{figure*}
\begin{center}
\includegraphics[width = 0.95\textwidth]{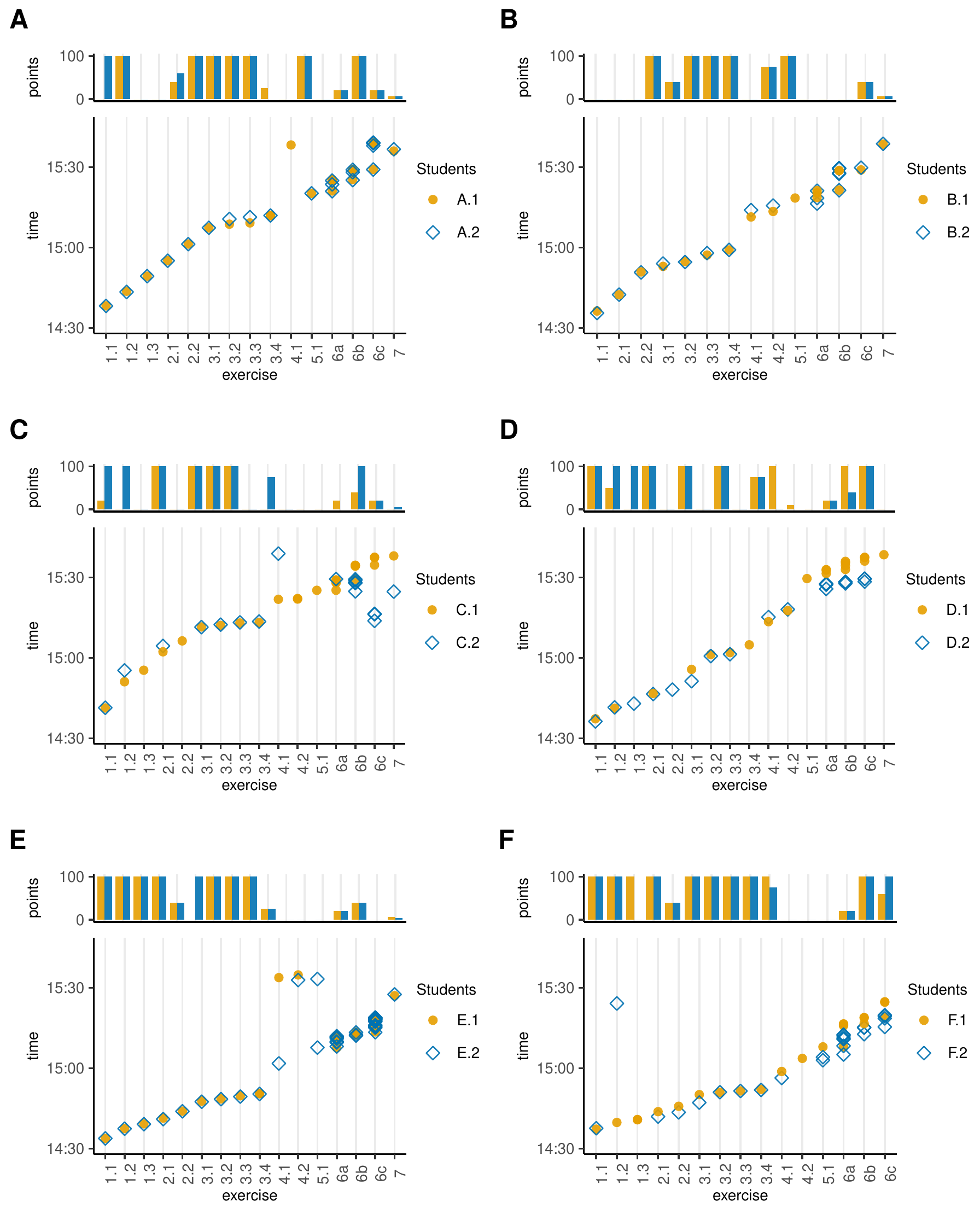}
\caption{Comparison of the event logs and achieved points for each of the test group's (2020/21) six lowest dissimilarity clusters (A - F).  The letters of the sub-figures correspond to the marked clusters in Figure~\ref{fig:tree_cl}. The number behind the letters refers to the node's left and right arm, respectively. At the bottom of each sub-figure, the sub-tasks are plotted against the clock time. Above the scatterplot, a bar chart is added to compare the points per subtask.}
\label{fig:comp_stud_20_21}
\end{center}
\end{figure*}

\Cref{fig:comp_stud_20_21} shows the exact course of events for the described selection of clusters. Each scatterplot plots all answers of the students in the cluster against their time of submission. We expect students' chronology to be more similar if their cluster's node is a lower height, i.e., lower dissimilarity. We also add the points achieved on top in a barchart.

As expected, all scatterplots show some kind of similarity. In particular, clusters A, B and E bear a striking resemblance. Their respective barchart further reveals these students to almost always achieve an equal number of points per subtask. In direct comparison, the plots of the remaining clusters C, D and F look less similar. This shows that clusters with lower node heights indeed contain more similar exams.

To assess whether these similarities are the result of collusion or coincidence, we repeat our approach on the comparison group, the final exam of the same course from two years ago. The plots were created analogously to the plots of the test group.

\begin{figure*}
	\includegraphics[width = \textwidth]{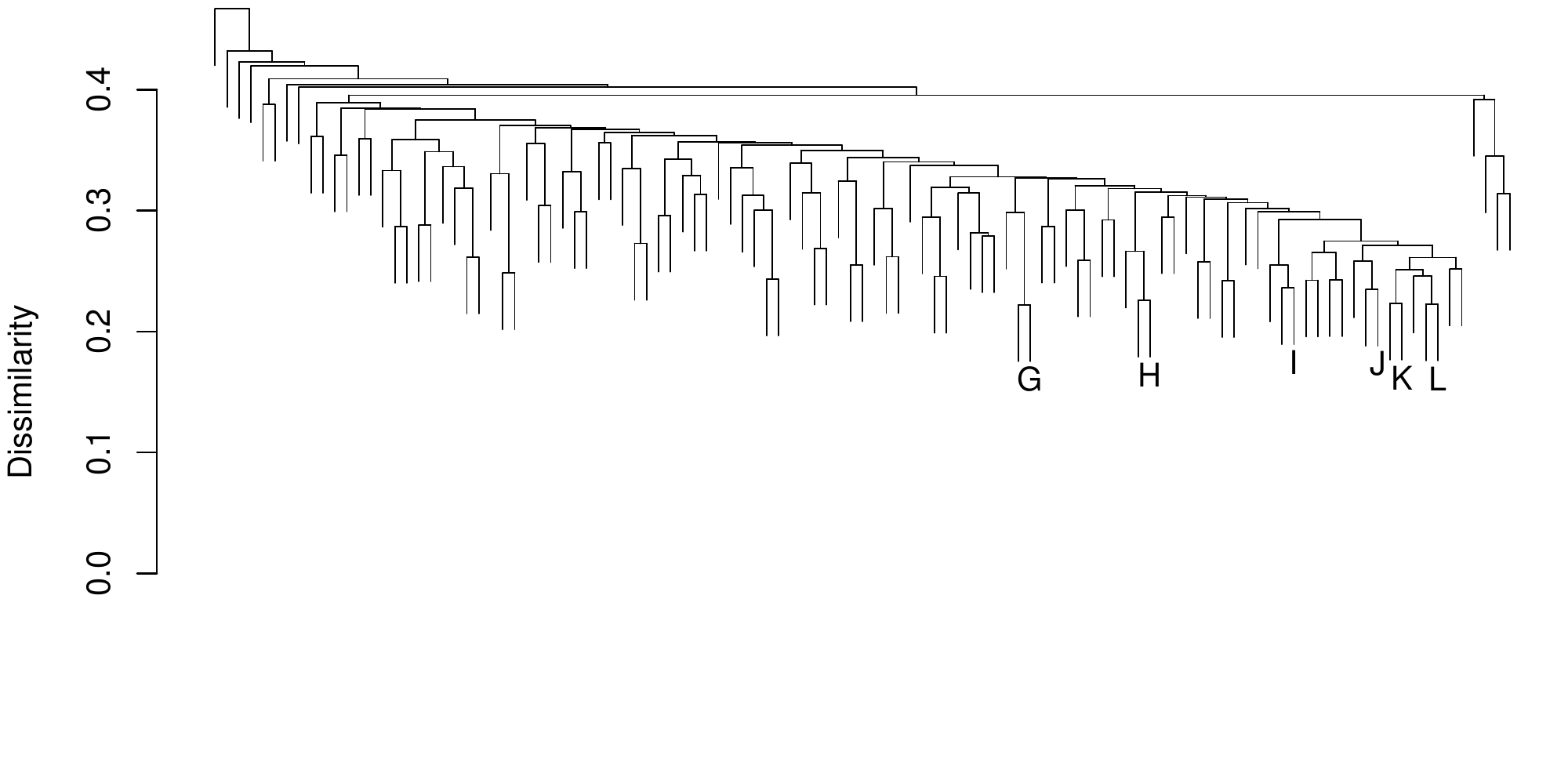}
	\caption{Dendrogram produced by average linkage clustering for the comparison group 2018/19. The dissimilarity of each group's node is displayed on the $y$-axis. Its value corresponds to the dissimilarity of the group's left and right member. G - L denotes the six lowest dissimilarity clusters.}
	\label{fig:tree_cl_18_19}
\end{figure*}

For the comparison group we also focus on average linkage clustering. The associated dendrogram, however, has a slightly different shape (see \Cref{fig:tree_cl_18_19}). The most prominent difference, in contrast to the test group, is the absence of any visually outstanding clusters. We rather observe most nodes at a comparatively similar height.

\begin{figure*}
	\begin{center}
		\includegraphics[width = 0.95\textwidth]{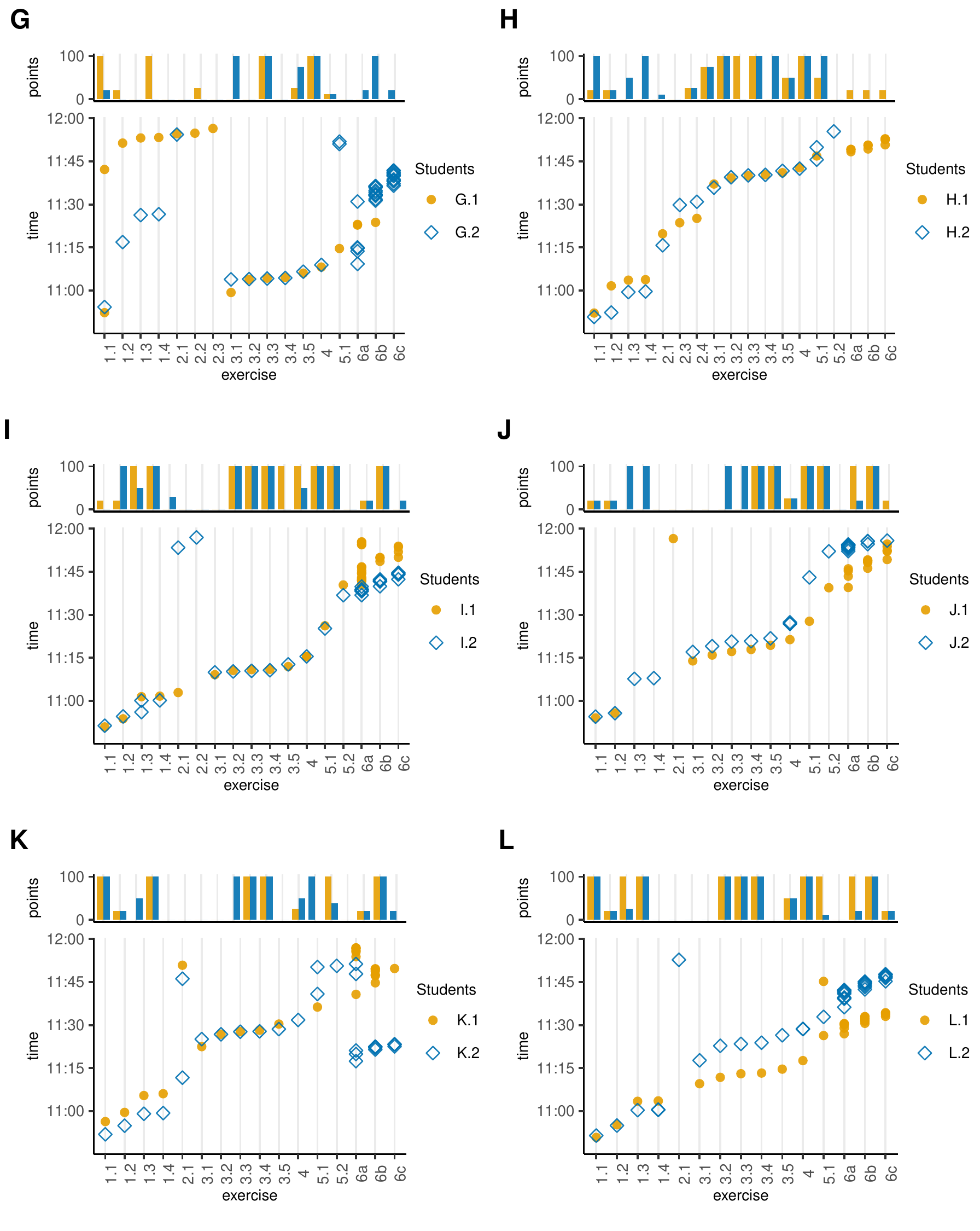}
		\caption{Comparison of the event logs and achieved points for each of the comparison group's (2018/19) six lowest dissimilarity clusters (G - L). The letters of the sub-figures correspond to the marked clusters in Figure~\ref{fig:tree_cl_18_19}. The number behind the letters refers to the node's left and right arm, respectively. At the bottom of each sub-figure, the sub-tasks are plotted against the clock time. Above the scatterplot, a bar chart is added to compare the points per subtask.}
		\label{fig:comp_stud_18_19}
	\end{center}
\end{figure*}

\Cref{fig:comp_stud_18_19} shows the scatter- and barplots of the six lowest clusters in the comparison group. We find there to be significantly less similarities not only in the chronological aspect, but also in the points achieved. 

The results from the comparison group support our assumption that widespread collusion over the entire exam is hardly achievable in presence. Moreover, the clear visual differences between comparison- and test group indicate that our findings in the latter might not be coincidental. 

In the test group it is relatively simple to identify at least three suspicious clusters. In the less obvious cases it can be challenging to decide which clusters to investigate further, as there exists no clear rule on where to draw the line between suspicious and unsuspicious cases. We address this issue by defining a ``normal degree'' of similarity, which can be used as a bound to classify whether a pair of students is deemed suspicious or not. For our data of the test group, there is no indication of the existence of suspicious clusters of more than two students. Hence, we refocus on the global pairwise dissimilarity $D(x_i, x_{i'})$. 

To assess the ``normal degree'' of similarity, we first standardise both distributions to improve their comparability. For the comparison group, we define a lower bound below which we categorise observations as extreme outliers. This bound is then used on the lower tail of the distribution of the test group. We want to identify cases in the unproctored test group which are rather extreme compared to the proctored comparison group. For this we calculate the lower bound as $Q_1 - 3 * IQR$ with $Q_1$ being the first quartile of the data and $IQR$ being the interquartile range. 

\begin{figure*}
\begin{center}
\subfloat[Non-Normalised]{\includegraphics[width = 0.5\textwidth]{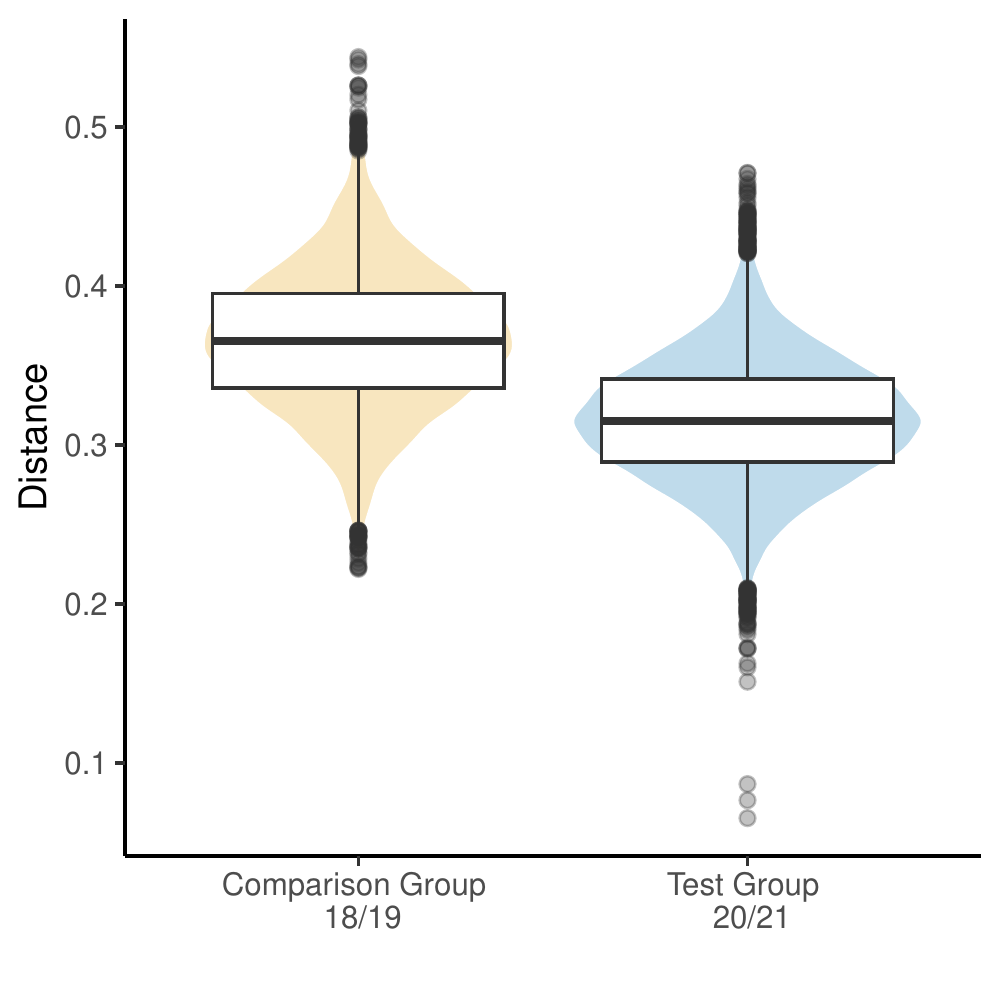}}
\subfloat[Normalised]{\includegraphics[width = 0.5\textwidth]{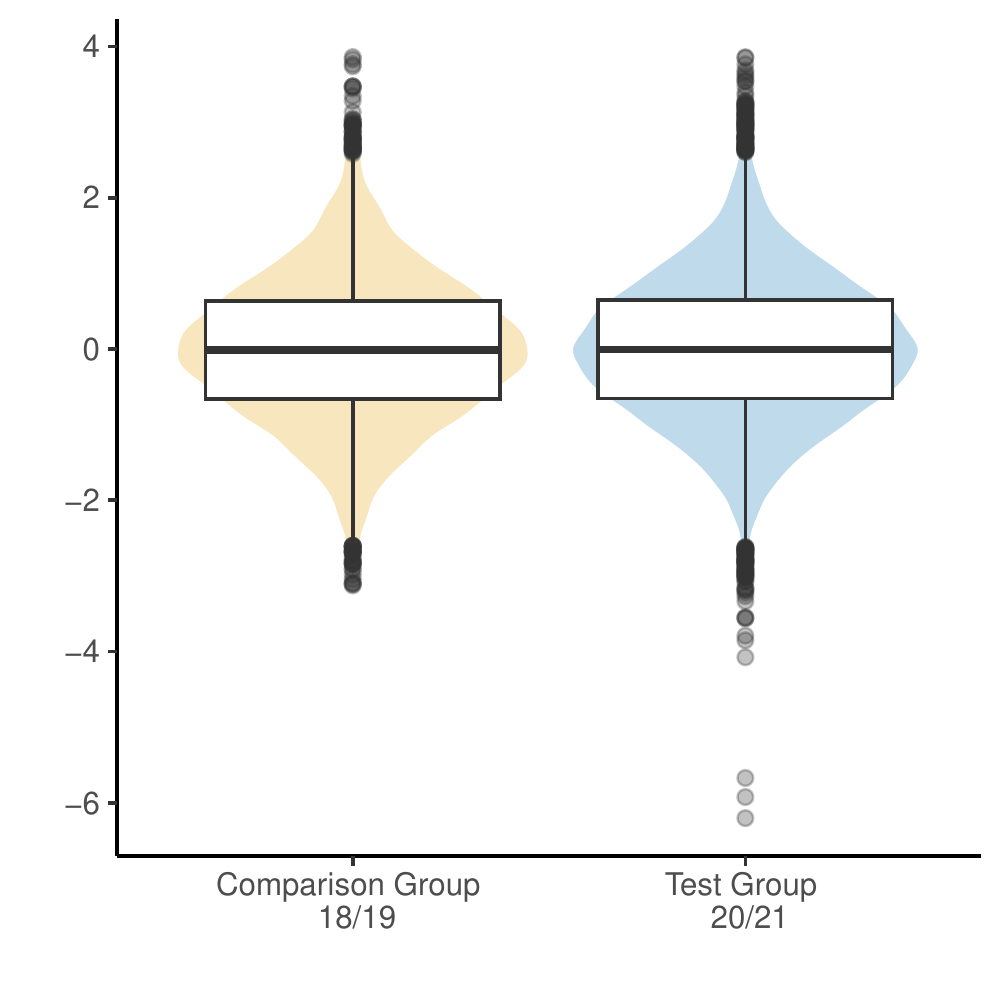}}
\caption{Boxplot of the pairwise distance measures.}
\label{fig:comp_stud_boxplot}
\end{center}
\end{figure*}

The boxplots in \Cref{fig:comp_stud_boxplot} show the distributions of the global pairwise dissimilarity $D(x_i, x_{i'})$ of all students in the comparison- and test group. A boxplot provides a graphic overview of location and dispersion of a distribution. The eponymous box marks the upper and lower quartile of the data. Outliers are displayed by individual points.

On the left hand side of \Cref{fig:comp_stud_boxplot} we observe that both distributions posess a similar shape, but a different median. The median value of the test group is significantly lower, indicating a lower average global pairwise dissimilarity in this group. Furthermore, we discover a high number of outliers in both groups, albeit at different positions in their respective distribution. In the test group, more outliers lie on the lower side of the box, with a greater distance to the main part of the distribution. We also find three observations with extremely small values on the lower tail of the test group's distribution. Unsurprisingly, these belong to the clusters A, B and E.

The right side of \Cref{fig:comp_stud_boxplot} shows the normalised distributions. It is apparent that the normalised distribution of the test group still contains more outliers. 

We apply the above mentioned lower bound on the test group's distribution to identify groups of students which are ``outstandingly similar''. Here, the before mentioned three cases (clusters A, B and E) fall below the lower bound for extreme outliers. While it is no surprise that these clusters were detected, our approach still aids us in deciding on when to stop inspecting further groups of students, as their level of similarity might as well occur in the comparison group.

To summarise, our approach offers a rule of thumb for narrowing down the number of suspicious cases. This is particularly useful if the visual distinction of cases is not clear-cut. 

\section{Conclusions and Discussion}\label{sec:conclusion}

During the COVID-19 pandemic many exams at universities had to be converted into unproctored take-home exams. We propose a method for detecting potentially colluding students in said exams. For this, we calculated a distance measure based on the students' event logs and their points achieved from the exam. Compared to former approaches adressing this topic, we use a distance measure which also applies if there exist multiple events per task. Subsequently, we use hierarchical clustering algorithms to detect clusters of potentially colluding students. The results show that our method is able to detect at least three clusters with near identical exams. 

To decide which degree of similarity might be more than a coincidence we compare the normalised distributions of the distance measures of our test and comparison group. We find pairs of students in the test group with values below the minimum of the comparison group. Thus, our approach provides a basis for deciding on which clusters are to be examined further. A limitation of this approach is that we do not know the ground truth in our groups and only be able to back up our reasoning on a comparison.

In summary, we have been successful in providing an opportunity to detect colluding students after the exam. We cannot say if this is sufficient evidence to initialise legal consequences. Nevertheless, we are confident that the higher chance of getting caught has a deterring effect on students. This would be an interesting direction for further research. Moreover, one could collect complementary evidence. By doing so, we found at least two of our suspicious students confirmed.

\section*{Acknowledgments}

Part of the work on this project was funded by the German Federal
Ministry of Education and Research under grant numbers 01PL16075 and 01JA1910 and by the Foundation for Innovation in University Teaching under grant number FBM2020-EA-1190-00081.

%
\bibliographystyle{abbrv}
\bibliography{bibliography}  
%
%

\newpage
\appendix
\section{Variable Description} \label{sec:vardesc}

\begin{table}[H]
\caption{Variable Description.}
\label{tab:vardesc}
\begin{center}
\begin{tabular}{cc}
\toprule
Variable & Description \\ 
\midrule
$s_{ij}$ & \parbox{6cm}{Points achieved in the $j$th subtask by the $i$th student.\\}\\ 
$v_{ij}$ & \parbox{6cm}{Even patterns for the $j$th subtask by the $i$th student.\\}\\ 
$x_{ij}$ & \parbox{6cm}{Measurement for the $i$th object and $j$th attribute. Here, $x_{ij}$ only functions as a variable for explaining the general clustering approach.\\}\\  
\bottomrule
\end{tabular} 
\end{center}
\end{table}

\section{Attribute Weights}  \label{sec:weights}

Below we list the attribute weights used for building the global object dissimilarity.

\begin{table}[h!]
{\footnotesize
\caption{The weights for all attributes in the test group (2020/21), rounded to three decimal places.}
\label{tab:weights2019}
\begin{center}
\begin{tabular}{ccc}
\toprule
& \multicolumn{2}{c}{weights} \\
\cline{2-3}
\rule{0pt}{12pt} (sub-)tasks & event patterns & points  \\
\midrule
\rule{0pt}{12pt}1.1 - 5.2 & 0.052 & 0.013 \\
\rule{0pt}{12pt}6a - 6c & 0.026 & 0.013 \\
\rule{0pt}{12pt}7 & 0.026 & 0.013 \\
\bottomrule
\end{tabular}
\end{center}
}
\end{table}

\begin{table}[h!]
\footnotesize{
\caption{The weights for all attributes in the comparison group (2018/19), rounded to three decimal places.}
\label{tab:weights1819}
\begin{center}
\begin{tabular}{ccc}
\toprule
& \multicolumn{2}{c}{weights} \\
\cline{2-3}
\rule{0pt}{12pt} (sub-)tasks & event patterns & points  \\
\midrule
\rule{0pt}{12pt}1.1 - 5.2 & 0.045 & 0.011 \\
\rule{0pt}{12pt}6a - 6c & 0.022 & 0.011 \\
\bottomrule
\end{tabular}
\end{center}
}
\end{table}

\FloatBarrier


\section{Dendrogram and Cophenetic Correlation Coefficient} \label{sec:dendro}

\begin{figure*}
	\begin{center}
			\subfloat[ Single Linkage Clustering]{\includegraphics[width = \textwidth]{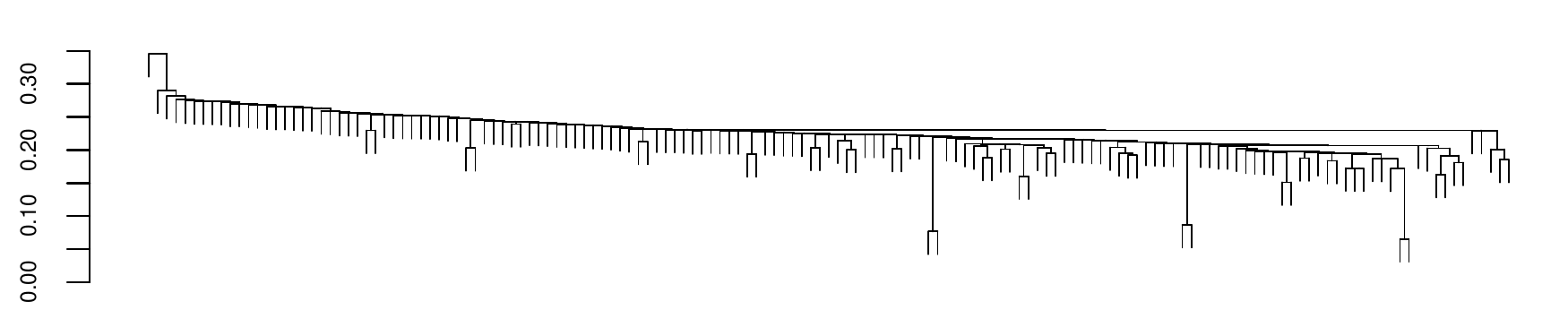}} \\
			\subfloat[ Complete Linkage Clustering]{\includegraphics[width = \textwidth]{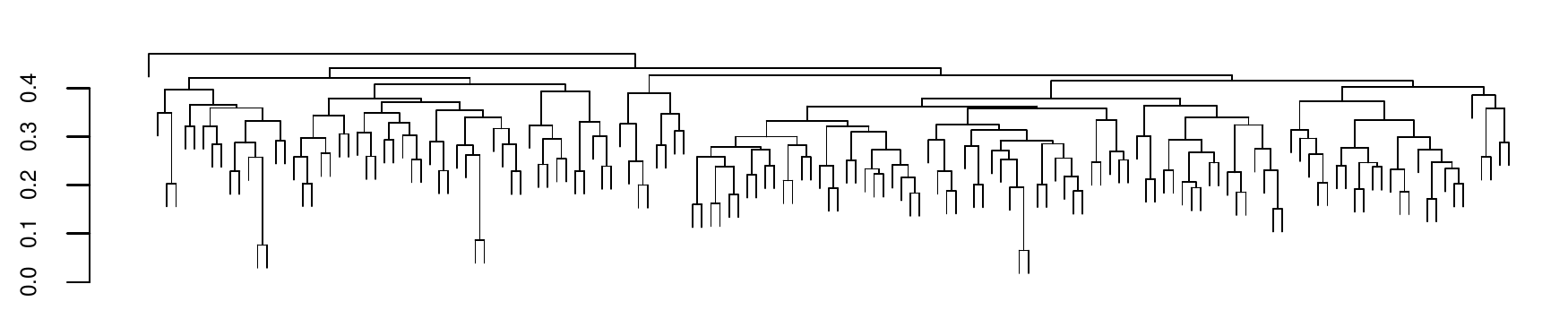}} \\
			\subfloat[Average Linkage Clustering]{\includegraphics[width = \textwidth]{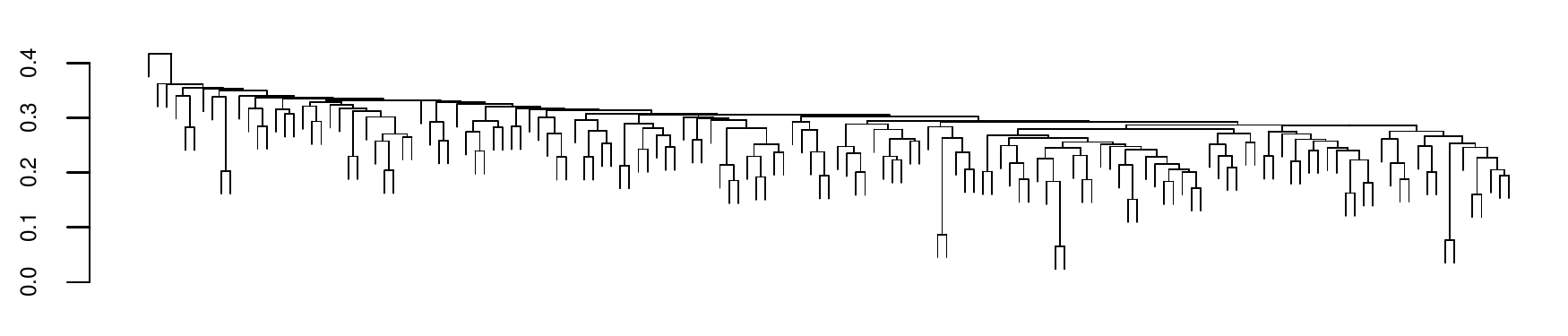}}
			\caption{Comparison of the dendrograms for all three linkage methods in the test group 2020/21.}
		\label{fig:dendro}
	\end{center}
\end{figure*}

\begin{figure*}
	\begin{center}
		\includegraphics[width = \textwidth]{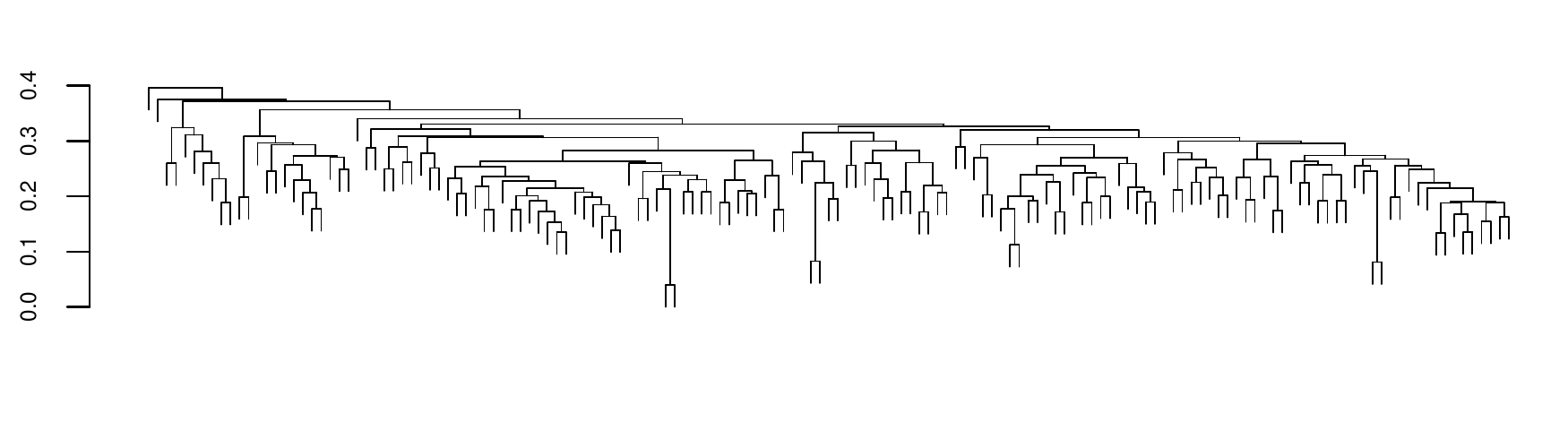}
		\caption{Dendrogram of average linkage clustering with the test data and equals weights.}
		\label{fig:dendro_rc}
	\end{center}
\end{figure*}

\begin{table}[h!]
	\centering
	\caption{The cophenetic correlation coefficient for all three linkage methods for the comparison (2018/19) and test (2020/21) group.}
	\label{tab:table_coph_cor_both}
		\begin{tabular}{@{}llc@{}}
			\toprule
			               & \multicolumn{2}{c}{C} \\ \cmidrule(l){2-3} 
			Linkage method & 2020/21   & 2019/18   \\ \midrule
			single         & 0.6610       & 0.6659     \\
			complete       & 0.4424      & 0.5051      \\
			average        & 0.6964      & 0.7595      \\ \bottomrule
		\end{tabular}%
\end{table}

 We must consider that clustering algorithms enforce a hierarchical structure on the data. This structure, however, does not have to exist. There exist a good amount of methods to assess how faithfully each algorithm represents the original distances in the data. Here, we use the cophenetic correlation coefficent ($C$). This is defined as the linear correlation between the pairwise dissimilarity $D(x_i, x_{i'})$ from the original distance matrix and the corresponding \textit{cophenetic} dissimilarity from the dendrogram $t(x_i, x_{i'})$, i.e., the height of the node of the cluster. Let $\overline{D}$ be the mean of $D(x_i, x_{i'})$ and $\overline{t}$ be the mean of $t(x_i, x_{i'})$. Then, $C$ can be written as

\begin{align}
C = \frac{\sum_{i<i'}\left(\left(D(x_i, x_{i'}) - \overline{D}\right)\left(t(x_i, x_{i'}) - \overline{t}\right)\right)}{\sqrt{\sum_{i<i'}\left(D(x_i, x_{i'}) - \overline{D}\right)^2 \sum_{i<i'}\left(t(x_i, x_{i'})-\overline{t}\right)^2}}.
\label{eq:coph_cor} \tag{B.1}
\end{align}

\Cref{tab:table_coph_cor_both} shows $C$ for all three linkage methods. The clustering with the complete linkage method seems to be the most unsuitable. The results of the single and average linkage clustering seem to be an adequate representation, with the latter a slightly better fit. We therefore proceed with the average linkage method in all further steps.

\section{Robustness Check} \label{sec:robust}
We repeat our analysis on the same data, but we assign the same weight to each attribute while calculating the global object dissimilarity matrix. In simple terms, we replace the weighted arithmetic mean in equation 3.1 in chapter 3.2 with the ordinary arithmetic mean.

\Cref{fig:dendro_rc} shows the dendrogram for the test data with equal weights. The algorithm still manages to identify the three suspicious clusters. Furthermore, these clusters are still merged at a comparatively low height.

\end{document}